\documentclass[sigconf,screen]{acmart}

\AtBeginDocument{%
  \providecommand\BibTeX{{%
    \normalfont B\kern-0.5em{\scshape i\kern-0.25em b}\kern-0.8em\TeX}}}

\setcopyright{acmcopyright}
\copyrightyear{2024}
\acmYear{2024}
\acmDOI{XXXXXXX.XXXXXXX}

\copyrightyear{2025}
\acmYear{2025}
\setcopyright{rightsretained}
\acmDOI{10.1145/3706598.3713286}
\acmISBN{979-8-4007-1394-1/25/04}
\acmConference[CHI '25]{CHI Conference on Human Factors in Computing Systems}{April 26--May 01, 2025}{Yokohama, Japan}
\acmBooktitle{CHI Conference on Human Factors in Computing Systems (CHI '25), April 26--May 01, 2025, Yokohama, Japan}

\usepackage{color}
\usepackage{soul}    
\usepackage{gensymb} 

\usepackage{graphicx}
 
\usepackage{enumitem} 

\usepackage{tabularray}

\newcommand{\systemName}{\textsc{Imprinto}}

\newcommand{\new}[1]{\textcolor{black}{#1}} 
\newcommand{\newCameraReady}[1]{\textcolor{black}{#1}} 

\begin{document}

\title[\systemName: Enhancing Infrared Inkjet Watermarking Human and Machine Perception]{\systemName:  Enhancing Infrared Inkjet Watermarking for \newline Human and Machine Perception}

\author{Martin Feick}
\orcid{0000-0001-5353-4290}
\affiliation{%
 \institution{DFKI \& Saarland University}
 \city{Saarland}
 \country{Germany}}
\affiliation{%
  \institution{MIT CSAIL}
  \city{Cambridge}
  \state{Massachusetts}
\country{US}
}
 \email{martin.feick@dfki.de}
 
\author{Xuxin Tang}
\orcid{0009-0004-7997-6280}
\affiliation{%
 \institution{Virginia Tech}
 \city{Blacksburg}
 \state{Virginia}
 \country{US}}
  \affiliation{%
  \institution{MIT CSAIL}
  \city{Cambridge}
  \state{Massachusetts}
\country{US}
}
\email{xuxintang@vt.edu}

\author{Raul Garcia-Martin}
\orcid{0000-0001-5319-2016}
\affiliation{%
  \institution{Universidad Carlos III de Madrid}
  \city{Leganes, Madrid}
  \country{Spain}
}
\affiliation{%
  \institution{MIT CSAIL}
  \city{Cambridge}
  \state{Massachusetts}
  \country{US}
}
\email{raulgarc@ing.uc3m.es}

\author{Alexandru Luchianov}
\orcid{0009-0005-9595-0635}
\affiliation{%
  \institution{MIT CSAIL}
  \city{Cambridge}
  \state{Massachusetts}
  \country{US}
}
\email{lknv@mit.edu}

\author{Roderick Wei Xiao Huang}
\orcid{0009-0001-7120-6968}
\affiliation{%
  \institution{MIT CSAIL}
  \city{Cambridge}
  \state{Massachusetts}
  \country{US}
}
\email{rwxhuang@mit.edu}

\author{Chang Xiao}
\orcid{0009-0008-7143-2771}
\affiliation{%
  \institution{Adobe Research}
  \city{San Jose}
  \state{California}
  \country{US}
}
\email{cxiao@adobe.com}

\author{Alexa Siu}
\orcid{0000-0002-4879-1476}
\affiliation{%
  \institution{Adobe Research}
  \city{San Jose}
  \state{California}
  \country{US}
}
\email{asiu@adobe.com}

\author{Mustafa Doga Dogan}
\orcid{0000-0003-3983-1955}
\affiliation{%
  \institution{Adobe Research}
  \city{Basel}
  \country{Switzerland}
}
\affiliation{%
  \institution{MIT CSAIL}
  \city{Cambridge}
  \state{Massachusetts}
  \country{US}
}
\email{doga@adobe.com}

\renewcommand{\shortauthors}{Feick, et al.}

\begin{abstract}

Hybrid paper interfaces leverage augmented reality to combine the desired tangibility of paper documents with the affordances of interactive digital media.
Typically, virtual content can be embedded through direct links (e.g., QR codes); however, this impacts the aesthetics of the paper print and limits the available visual content space. 
To address this problem, we present \systemName, an infrared inkjet watermarking technique that allows for invisible content embeddings only by using off-the-shelf IR inks and a camera.
\systemName~ was established through a psychophysical experiment, studying how much IR ink can be used while remaining invisible to users regardless of background color.
We demonstrate that we can detect invisible IR content through our machine learning pipeline, and we developed an authoring tool that optimizes the amount of IR ink on the color regions of an input document for machine and human detectability. 
Finally, we demonstrate several applications, including augmenting paper documents and objects. 
\end{abstract}




\begin{teaserfigure}
  \includegraphics[width=\textwidth]{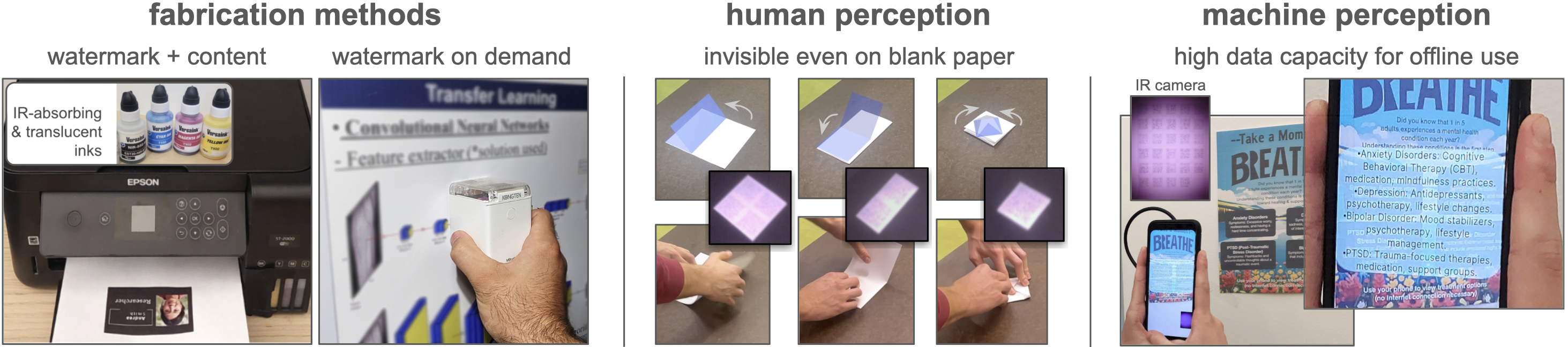}
  \caption{\systemName~ are real-life documents that enable watermarking through the use of IR-absorbing ink. Users can apply it by (1) making it part of the document's fabrication process using an off-the-shelf inkjet printer or (2) by using a handheld inkjet printer on demand. The embedded content is invisible to users, even on blank paper, while allowing for high data capacity. Our system reliably decodes the IR content and \textit{is further instructed by it} regarding what images to track and what to display.}
  \label{fig:teaser}
\end{teaserfigure}

\maketitle

\section{Introduction}

Despite the growing adoption of digital devices, the use of paper documents remains prevalent in our daily lives~\cite{gitelman_paper_2014}.
However, paper-based information can often lack the interactivity and accessibility of digital media~\cite{holman_paper_2005}.
In recent years, various paper augmentation techniques have been proposed to bridge this gap, enabling users to enhance the information content of their documents by seamlessly integrating digital media~\cite{han_hybrid_2021, li_holodoc_2019, rajaram_paper_2022}.
However, the main disadvantage of these techniques is that the printed markers interfere with printed paper media. 
As a result, they can affect the aesthetics of the print, breaking the illusion that digital is seamlessly embedded in the physical.
This goes as far as distracting consumers from the relevant parts or even degrading overall readability and accessibility.  
This has prevented techniques from going beyond single links or QR codes hidden at the bottom of the document to seamless paper document augmentations with rich digital content~\cite{dogan_standarone_2023}.

To address this challenge, we present \systemName\footnote{The system name \systemName~ is a combination of \textit{imprint} and \textit{pronto} (quickly), indicating that existing documents can be quickly watermarked for additional AR content or metadata embedding.}, a novel infrared inkjet watermarking technique that allows for invisible paper augmentation.
\systemName~ enables users to interact with printed documents in new ways by embedding hidden high-capacity digital information that can be accessed using a smartphone equipped with an infrared (IR) sensor. The technique works by printing an invisible watermark using an IR-absorbing ink. When the paper is illuminated with IR light, the watermark becomes detectable and can be captured by an IR camera.
In contrast to previous work, \textbf{our watermarking approach leverages the whole document area, including text, graphics, and white space, to increase capacity and support the embedding of a larger amount of information while remaining invisible to the naked eye.}
By utilizing commercially available inkjet printers and inks, we rapidly fabricate these encoded documents without having to include complicated or manual processes such as spraying \cite{willis_hideout_2013} or screen printing \cite{kim_ministudio_2016}.

To inform our system, we first conducted a psychophysical experiment to determine estimates for the maximum amount of IR ink we can embed in paper documents while preventing the IR watermarks from becoming visible to users.
Our results show that IR ink can remain invisible to the human eye, regardless of the background color, i.e., even on white paper. 
Next, we demonstrate that invisible document watermarks can be captured and decoded using our mobile IR sensor and a machine learning (ML) pipeline based on a convolutional neural network (CNN), which binarizes IR watermarks into decodable black-and-white patterns.
Our technical evaluation investigates various factors related to the IR ink machine detectability and data capacity, providing guidelines for content creators and designers who want to use our invisible watermarking technique.  
Together, our results demonstrate that we can embed invisible content into paper documents that can still be decoded by an off-the-shelf hardware and software stack.

To facilitate the embedding process, we developed a software tool that uses the determined IR ink values to embed digital content invisibly onto paper documents in the form of QR codes. 
Here, a user simply uploads an image, specifies the number of watermark embeddings, and marks where these should be located. The software analyzes the background colors given the provided embeddings and paper regions, recommending the amount of IR ink that can be used while remaining invisible. The user can then directly fabricate the document using conventional inkjet printing (see \autoref{fig:teaser}). 

We demonstrate the versatility of \systemName~ through a series of applications, which highlight a variety of use cases covering education, arts and design, desktop publishing, robotics, entertainment, sustainability and security.
Unlike other paper-based augmentation techniques, \new{\textbf{\systemName~ is invisible, works with new and existing documents, and does not require any embedded electronics or special paper.}}

\vspace{0.2cm}
To summarize, the contributions we make are as follows:
\begin{itemize}[leftmargin=0.5cm]
    \item  \systemName, an inkjet printing-based watermarking method that allows the utilization of the whole sheet (both blank and color-printed areas) for invisible data embedding onto new and existing physical media.
    \item A methodology on how to evaluate human perception of IR inkjet ink, and its optimization for invisibility, data capacity and machine detection.
    \item A universal hardware detection module compatible with \textit{USB-C} devices.
    \item An open-source\footnote{\url{https://imprinto.github.io}} machine learning pipeline for binarizing the captured IR watermarks for robust detection and data capacity under different visible background content and IR ink levels.
    \item A set of use cases that demonstrate  \systemName's application space, even beyond paper-based documents.
\end{itemize}
\section{Related Work}

Combining the digital and the physical worlds has been one of the main uses cases of augmented reality~\cite{dogan_fabricate_2022, dogan_augmented_2024}. 
Below we explain how the two worlds interact with each other, how paper specifically is used for these interactions, 
and how tags are used to specify interactions in previous works.

\subsection{Hybrid Paper-Digital Interfaces}
The benefits of hybrid paper-digital interfaces have been extensively explored in prior work~\cite{rajaram_paper_2022, li_holodoc_2019, alessandrini_audio-augmented_2014, song_penlight_2009}. Han et al. conducted a systematic literature review and defined hybrid paper-digital interfaces as \textit{"any interface embedding digital or electronic functionality in physical paper to enable its use as an input or output device"}~\cite{han_hybrid_2021}.

To enable input and enhance digital paper applications, researchers proposed integrating electronics into paper sheets, such as sensors~\cite{gong_printsense_2014, jacoby_drawing_2013}, actuators~\cite{qi_electronic_2010}, or illuminated displays~\cite{klamka_illuminated_2017, klamka_illumipaper_2017}.
Buechley et al. further proposed \textit{paper computing}~\cite{buechley_paints_2009} as a concept to allow users to create functional computational artifacts on painted paper substrates using a combination of these paper-embedded components as well as microcontrollers.
On the other hand, leveraging AR  is a common technique used to add a digital information layer but avoid integrating electronics into paper.
For instance, \textit{Dually Noted}~\cite{qian_dually_2022} and \textit{HoloDoc}~\cite{li_holodoc_2019} demonstrate the utility of augmenting paper for improving productivity tasks for phone- or headset-based AR.
\textit{PapARVis}~\cite{chen_augmenting_2020} and \textit{Paper Trail}~\cite{rajaram_paper_2022} explore authoring workflows that support educational and data visualization use cases.
\textit{PaperLens}~\cite{spindler_paperlens_2009} tracks sheets in 3D space above a top-projected tabletop~\cite{steimle_physical_2010} to visualize and interact with 3D information spaces.
\textit{SpARklingPaper}~\cite{drey_sparklingpaper_2022} augments paper sheets from below using a tablet’s screen placed underneath.
The paper augmentation in these projects are enabled by systems such as dedicated standalone apps that were tailored for specific use cases. Most of these leverage image tracking libraries such as \textit{Vuforia}\footnote{\url{https://developer.vuforia.com/}}, \textit{ARKit}\footnote{\url{https://developer.apple.com/augmented-reality/arkit/}}, and \textit{ARCore}\footnote{\url{https://developers.google.com/ar/}}. These utilize visual features (\textit{SURF}, \textit{SIFT}) or fiducial markers (\textit{ArUco}~\cite{garrido2014automatic}, QR codes\footnote{\url{https://www.qrcode.com/en/}}, \textit{ARToolKit}~\cite{kato1999marker}),  which is based on previous research in computer vision (CV)~\cite{wagner2010RealTimeDetection, Wagner2008PoseTracking}. While visual feature based tracking does not require the addition of new visual elements such as QR codes, they do not carry high capacity to be able to represent a \textit{\textbf{unique and direct} anchor 
to digital experiences}.
In contrast, \textit{AniCode}~\cite{wang_anicode_2019} uses QR codes to store entire animation sequences contextual to an object, without having to access digital content from an external service.
A remaining limitation is that 
the visible QR codes become obtrusive in the user's view as longer sequences are embedded.

In this work, we investigate invisible IR watermarking as the fabrication method to enable hybrid documents to be portable containers of \textit{both} physical (visual) and digital (invisible) content. Our aim is to preserve documents' defining attribute of portability by \textit{making its digital link \textbf{an intrinsic part} of the document}.

\subsection{Linking Digital Information to Physical}

To connect paper documents to digital content, machine-readable printed codes, such as barcodes, have been actively used as a compact and cheaper way to embed digital information~\cite{dogan_ubiquitous_2024}. 
2D barcodes, such as QR codes, store information in the form of contrasting bits, but affect the original design as they occupy space and impact the visual look of the final artifact~\cite{getschmann_seedmarkers_2021, song_my_2018}, particularly those that store large amounts of data~\cite{denso_wave_information_2022}. 
Thus, researchers explored making more unobtrusive camera-detectable tags to create a more seamless, discreet user experience~\cite{dogan_g-id_2020, dogan_structcode_2023}.
For this goal, works investigated subtle pixel manipulations to encode data \textit{unobtrusively}~\cite{fu_chartem_2021, xiao_fontcode_2018, tancik_stegastamp_2020, zhang_viscode_2021}.
For instance, \textit{Chartem}~\cite{fu_chartem_2021} encodes subtle pixels into chart background, 
however, it only supports digital versions of charts, i.e., the data cannot be robustly decoded once they are printed.
Methods that support printed documents can only leverage colored pixels and are thus limited in the amount of stored information. 
For instance, \textit{StegaStamp}~\cite{tancik_stegastamp_2020} can encode up to 7 characters into a single image, and \textit{FontCode}~\cite{xiao_fontcode_2018}  encodes 1.77 bits per character (e.g., a 1,000-character paper abstract would store 221 encoded characters).
Furthermore, they cannot store information in non-printed white space, which is a typical part of documents.

\systemName~leverages the \textit{\textbf{whole} document area, including text, graphics, and white space, to increase capacity and support the embedding of \textbf{digital (invisible)} content regardless of the physical (visual) content layout.}

\subsection{Aesthetically Integrated Visual Markers}
Another relevant approach is to make the markers human-designable so that they seamlessly blend into the aesthetics of the content.
This relies on visual features that follow a predefined set of rules, allowing the system to decode the integrated marker. 
For example, \textit{d-touch} \cite{Enrico_visualMarkers_2009} uses the relationship of dark and light regions in the image for binary encoding.
While the initial technique was limited to smaller areas and black and white, \citet{Preston_MarkerScale_2017} extend this approach to impressive scale and aesthetics by adding area order codes and visual checksums.
\citet{Benford_Decoration_2017} went beyond paper-based interfaces by embedding computer-readable codes in the form of decorations to ceramic bowls, fabric souvenirs, and an acoustic guitar, demonstrating the wide application space.

Studies showed that designers can incorporate these high-capacity codings in their visual designs \cite{Enrico_visualMarkers_2009, Jung_HumanDesignable_2019}.
However, it may require additional mental effort, be prone to errors, and limit artistic freedom.
They are further challenged by the necessity to adapt the visual design to specific rules and remain limited to areas displaying content. Moreover, visual markers can only be used during the creation process \cite{Jung_HumanMarker_2019}, whereas \systemName~\textit{can be used as \textbf{part of the fabrication process} or added \textbf{on demand}.}
Nevertheless, this impressive line of research demonstrates how high-capacity markers can seamlessly blend into artistic images, inspiring us to look into alternative approaches that allow invisible data embeddings while preserving the aesthetics of new and existing content.
Ultimately, we hope the techniques could complement one another to allow for even higher data capacities.

\begin{figure*}[]
  \centering
  \includegraphics[width=0.85\linewidth]{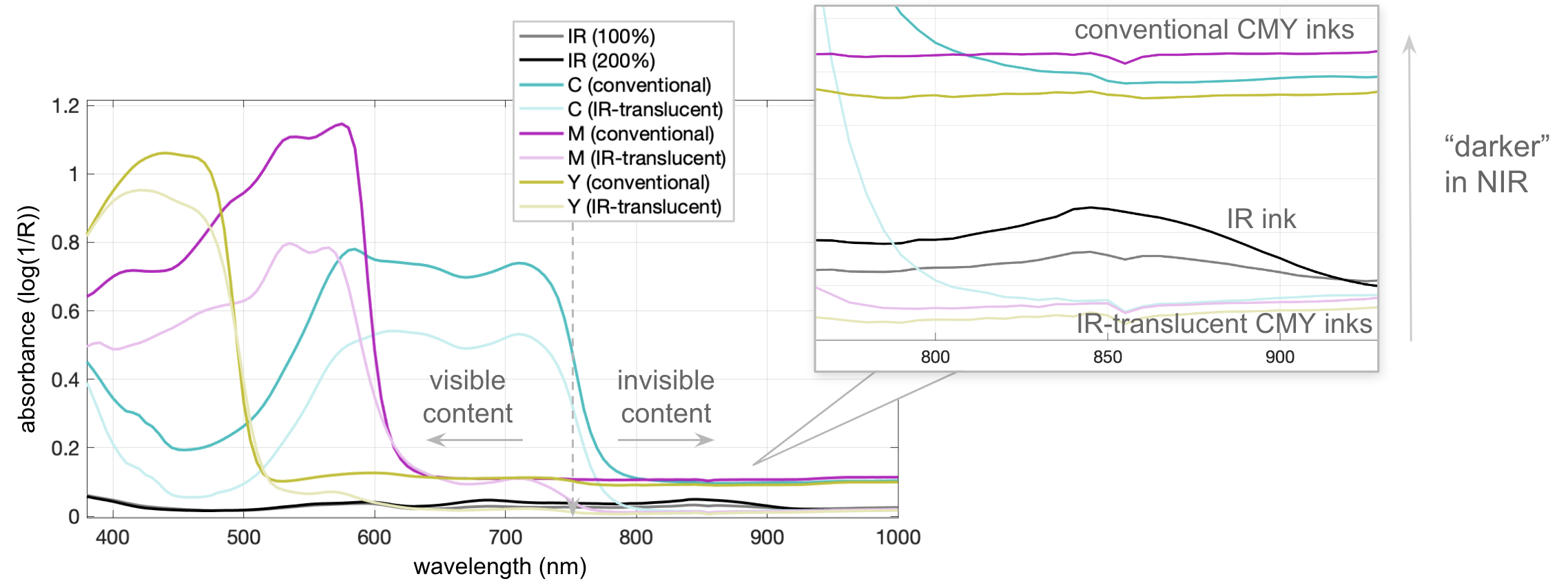}
  \caption{Spectral analysis of the inkjet inks.}
  \Description{.}
  \label{fig:inkSpectrumFigure}
\end{figure*}

\subsection{Infrared-Based Tags}
\new{IR based approaches such as patterns projected by external sources~\cite{lee_hybrid_2007, Kakehi2006Transparent}, materials that pass~\cite{dogan_infraredtags_2022}, reflect~\cite{kim_ministudio_2016, balaji_retrosphere_2023}, absorb~\cite{rosner_spyn_2010, willis_hideout_2013}, or fluoresce~\cite{dogan_brightmarker_2023} IR light to embed patterns have emerged.}
For instance, 
\textit{MiniStudio}~\cite{kim_ministudio_2016} tracks objects by adding IR-reflective stickers
with fiducial markers
using screen printing.
\textit{InfraredTags}~\cite{dogan_infraredtags_2022} covers fiducial markers with an opaque material that passes IR light to make them imperceptible.

As for IR-absorbing materials, 
\textit{Spyn}~\cite{rosner_spyn_2010} adds such inks into knitted artifacts to embed hidden information.
\textit{HideOut}~\cite{willis_hideout_2013} creates hidden tracking markers from IR-absorbing ink to project digital imagery on physical paper; however, a spray gun is needed to evenly coat the paper surface.
Wang et al.~\cite{wang_design_2008} showed how different IR absorption characteristics of CMY and black ink can be used to subtly embed fiducial markers; however, it is limited to dark areas\footnote{\url{https://frauzufall.de/en/2020/print-invisible-ir-markers-using-a-standard-printer/}}.
Similar to this work, we leverage the convenience and accessibility of commodity inkjet printers, commonly used in ubiquitous computing research~\cite{kawahara_instant_2013,cheng_silver_2020}, which allow us to print hybrid interfaces rapidly using the encoded data, but \systemName~\textit{also makes use of the \textbf{empty space} in the document}.

\vspace{0.2cm}
To summarize, 
\systemName~ has three main benefits that allow it to rapidly create portable 
hybrid document experiences:
(1) We utilize the whole document area, including text, graphics, and white space to increase the freedom of placement and amount of data that can be invisibly stored in new and exciting content. 
(2) We can rapidly create these encoded documents without having to use complicated or manual processes such as spraying or screen printing, thanks to commercially available inkjet printers and inks.
(3) We build on previous demonstrations~\cite{dogan_standarone_2023} and establish a scientific basis on how to use perception research to truly hide digital IR content for different classes of visibly printed content, while optimizing for machine detection and information capacity using ML.

\section{Infrared-Absorbing and -Translucent Inkjet Inks}
\label{inks}

One key component in our system is the set of inks used in our printed samples.
For our implementation, we used an off-the-shelf desktop inkjet printer (\textit{Epson ST-200}), \new{matte paper from \textit{Canon}\footnote{\new{\url{https://www.amazon.com/Canon-7981A004-Photo-Paper-Sheets/dp/B0000721Z3/}}}}, and four types of inkjet-compatible inks (C, M, Y, IR) by \textit{VersaCheck}\footnote{\url{https://www.versacheck.com/}}.
The desktop printer has four refillable tanks, which we leveraged for visible and invisible content printing. The cyan, magenta, yellow (CMY) channels of the printer were filled with the infrared-translucent C, M, Y inks, and the $K$ (i.e., conventionally black) channel was filled with the IR ink (NIR, $78~ml$ bottle). This ink passes light at visible wavelengths, and thus appears mostly transparent to the naked eye, but has high absorbance intensity in the near-infrared (NIR) range, with the peak absorbance wavelength at $850~nm$. Thus, it is captured as black under an infrared camera lens. Alternately, users can leverage the flexibility of handheld printers \cite{Pourjafarian2022Print} such as the \textit{Kongten MBrush}\footnote{\url{https://www.amazon.de/-/en/EVEBOT-Kongten-Mbrush-Printer-Android/dp/B0C84P61WH?th=1}} to embed watermarks on demand, only requiring IR ink. However, we used the desktop printer for our investigations of human and machine perception as it allowed us to produce very consistent and reproducible results.

As the ink loaded into the "black" channel is non-black (for human sight), we instead generate the black visible content by mixing the C, M, and Y inks, as their combination results in black color.
\new{We achieve this elimination of the black channel requirement by converting the original RGB color space of the visual part of the document to CMY (i.e., as opposed to CMYK) by setting the $K$ channel to zero through Gray Component Replacement (GCR)~\cite{kang_gray_1994} via \textit{Adobe Photoshop}. We note that the resulting black color may slightly visually differ from off-the-shelf black ink.}
The visible color content of the document is printed this way initially, then the sheet is fed back to overlay the invisible content with the NIR ink. The watermark's content is then printed using the $K$ channel only.

\begin{figure*}[]
  \centering
  \includegraphics[width=0.9\linewidth]{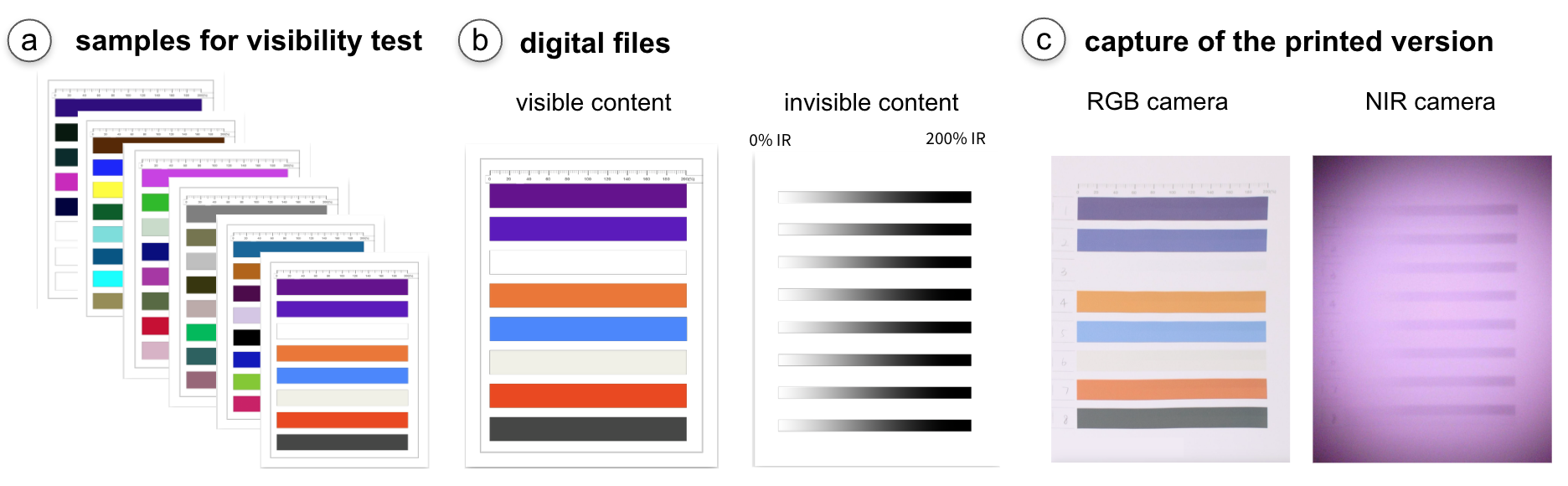}
  \caption{User study samples.}
  \Description{.}
  \label{fig:userStudyFigure}
\end{figure*}

\subsection{Absorbance Analysis of Inks}
\label{spectrum}
We conducted a spectral analysis on the four types of inks we use in our system (infrared-translucent C, M, Y, and infrared-absorbing IR)  with the help of a spectrophotometer, the \textit{Perkin Elmer Lambda 1050 UV/VIS/NIR}\footnote{\url{https://www.perkinelmer.com/product/lambda-1050-2d-base-inst-no-sw-l6020055}}.
In addition, we also measured the properties of \textit{conventional} CMY inks, so we can compare it them to our \textit{infrared-translucent} CMY inks. Our results are shown in \autoref{fig:inkSpectrumFigure}.
The x-axis represents wavelength of light measured in nanometers ($nm$), and its range $380$-$1100~nm$ was selected to illustrate the behaviour under visible and NIR light conditions.
The y-axis is log(1/R), which indicates the absorbance rate calculated via the reflection (R) measured by the device.

Within the human visibility range (from $380~nm$ to $750~nm$), both regular and infrared-translucent CMY inks are more discernible to the human eye compared to the IR ink, which registers close to zero visibility and is therefore harder for humans to perceive. However, beyond $750~nm$, the absorbance of conventional CMY inks drop to around $0.1$, while those for infrared-translucent CMY inks drop closer to $0$. At around $800~nm$, the absorbance of the IR ink surpasses that of the three infrared-translucent CMY inks, as shown in the callout of \autoref{fig:inkSpectrumFigure}. 
This means that starting at this wavelength, the IR ink would look "dark" (i.e., absorbing) to an IR camera in comparison to translucent CMY inks.
Notably, at $850~nm$, the difference between the IR ink and infrared-translucent CMY inks is the highest.
The larger this difference, the easier it is to distinguish the watermark from the the CMY content under an IR camera.
Thus, at this wavelength, one could more reliably decode the data printed with IR ink, as the CMY content would look most dispersed.
Thus, we use IR ink in combination with these IR-translucent CMY inks.
In comparison, conventional CMY inks are even "darker" than IR ink at this wavelength, meaning they would block the IR content from being detected if they were used for printing the documents instead.

Consequently, IR cameras should be calibrated to image at $850~nm$ with the background printed using our infrared-translucent CMY inks to enhance the detectability of the IR ink.

\vspace{0.3cm}

\section{Experiment: Detectability of IR Ink}
\label{visibility-experiment}

One challenge with IR ink is that, even though it allows for invisible pattern embedding, it may become visible to the naked eye when its concentration is too high. This results from a difference in intensity, that is, a contrast between the IR-inked parts and the rest of the paper region. Consequently, it can potentially affect the readability and aesthetics of the print. Our central interest lies in using the highest concentration of IK ink to improve the detection rate while remaining invisible to human eyes. \new{The experiment was conducted indoors under normal office lighting conditions (i.e., well-lit, about 300 lux and a color temperature of 3500K–-5500K)}.

To investigate this, we conducted a psychophysical detection threshold experiment on the detectability of printed IR ink at different background colors. We included colors with varying RGB values, darkness ($K$)~\cite{li_black_2013} and of different luminescence~\cite{ridpath_techniques_2000} to cover a large color spectrum and to understand how these color properties affect the detectability. This decision was motivated through multiple feedback loops and informal pilot testing, suggesting, e.g., that darker colors mask the IR ink, whereas IR ink on brighter backgrounds can be more easily noticed by humans. To improve the quality and robustness of the data collection, we ran the experiment in the lab under controlled lighting conditions. 

We formulated the following two hypotheses in our experiment:

\textbf{(H.1)} Inkjet printed IR ink can remain invisible to human eyes, regardless of background color.

\textbf{(H.2)} The maximum density of IR ink that remains invisible depends on the properties of the background color.

\subsection{Method}
To determine estimates for detection thresholds, we used the psychophysical methods of limits \cite{kingdom_chapter_2016}, exposing participants to ascending stimuli, that is, increasing the concentration of IR ink, until they noticed it. Participants were presented with two color bars, one with IR ink and the other without IR ink, and we asked them to compare both color bars and report if they noticed a difference between them using a 2-alternative-forced-choice (2AFC) method. As depicted in \autoref{fig:userStudyFigure}, we used a gradient pattern of IR ink, starting with no IR ink 0\% (left) to 200\% (right) density. By using 0\% as the starting stimuli, we ensure that participants start with a level of IR ink that cannot be noticed \cite{leek_adaptive_2001}. The higher IR ink densities of 101\% to 200\% require two successive printing procedures. Given the novelty of our approach, we determined a reasonable range of IR ink densities through pilot testing. 

\begin{figure}[]
  \centering
  \includegraphics[width=0.45\textwidth]{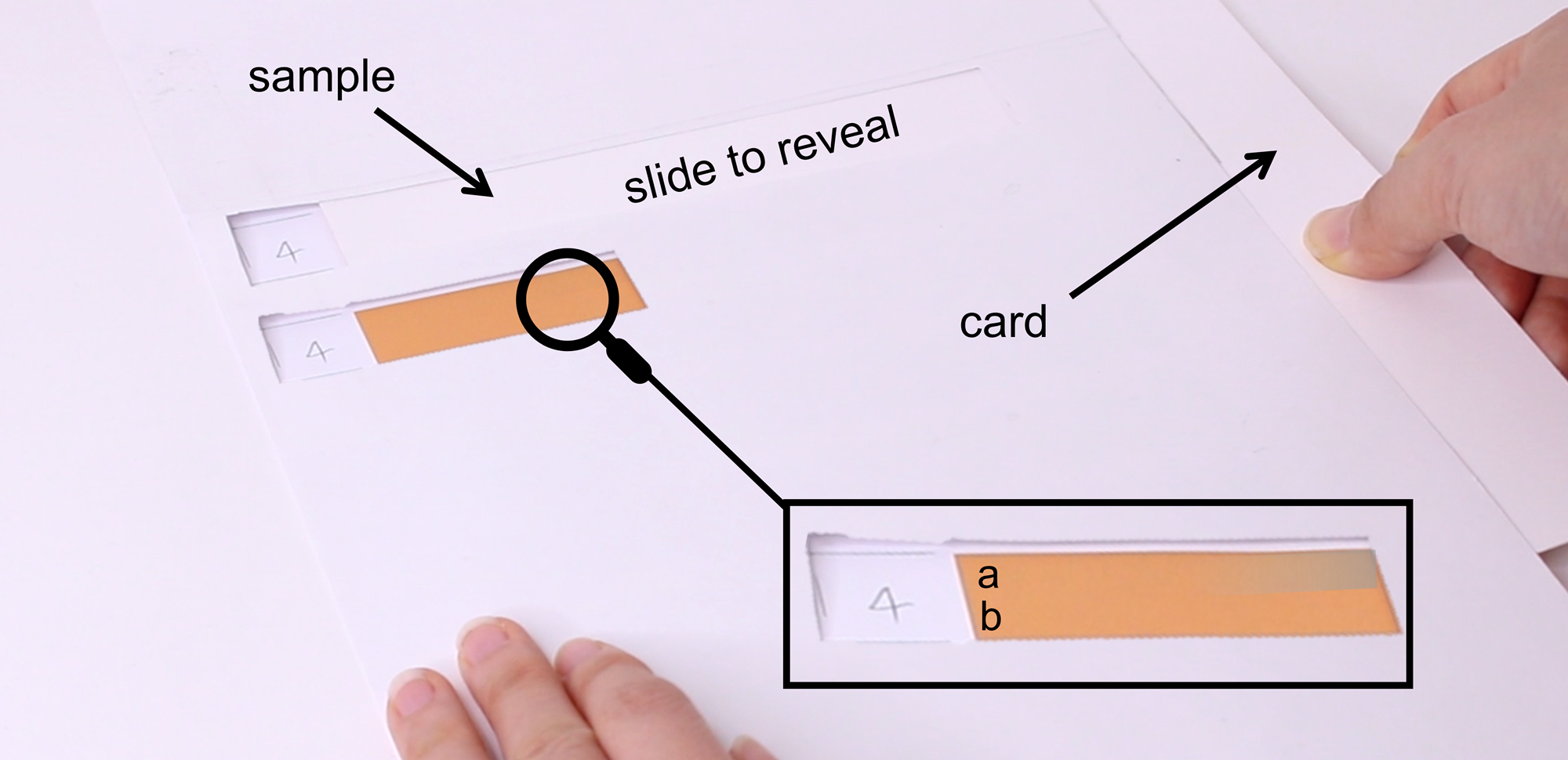}
  \caption{Study procedure. The participant slides the card steadily to the right to reveal the color bar. When participants noticed a color difference between the top (a) and the bottom (b) of the color bar, they were told to report it.}
  \Description{.}
  \label{fig:task}
\end{figure}

\paragraph{Task.} Participants moved a thick white paper card (depicted in \autoref{fig:task}) from the left (starting stimuli) to the right. We asked participants to move the card slowly and evenly, stopping and reporting immediately when they detected the IR ink. The experimenter recorded the position at which they stopped using the ruler printed at the top of the page (see \autoref{fig:userStudyFigure}), which was not visible to the participants. Consequently, participants continued with the next row until they completed all rows. Only one row was visible to the participants at a time, and everything else on the page was covered. Participants were informed about the procedure, and we explicitly showed them IR patterns multiple times during the warm-up phase to ensure that they were able to detect it. We asked them to report the IR ink as soon as they noticed it, thus targeting the most conservative case. Participants took breaks whenever needed.

\begin{figure*}[h]
  \centering
  \includegraphics[width=\linewidth]{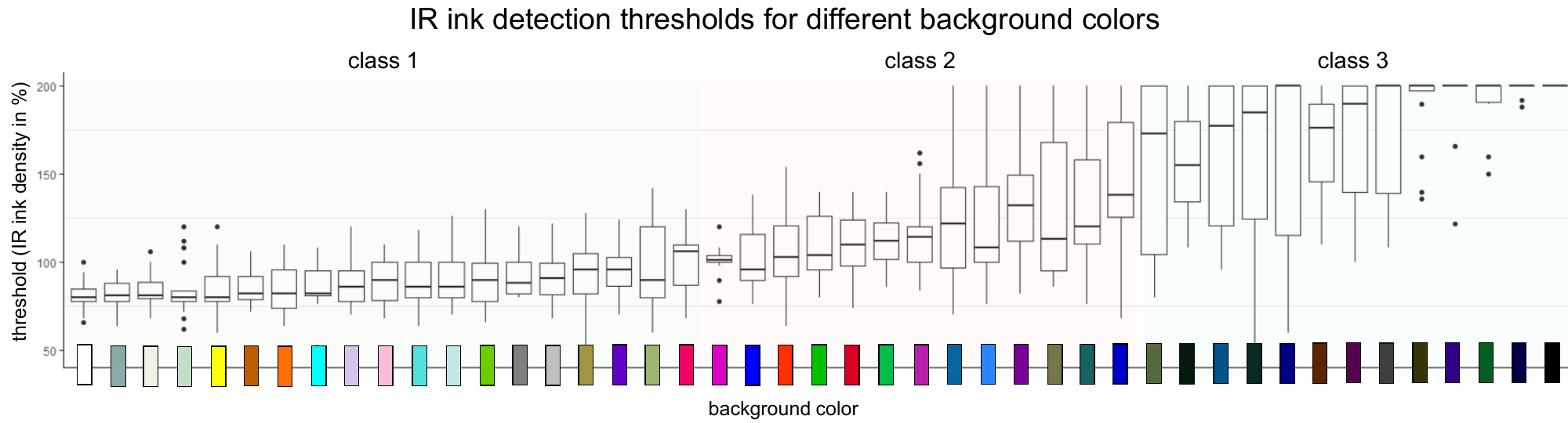}
  \caption{IR ink DTs for each color tested in our experiment, reaching from a mean DT of 81\% all the way to 200\% IR ink density. \new{The three classes are provided based on participants' ability to detect the IR ink.}}
  \Description{.}
  \label{fig:thresholds}
\end{figure*}

\subsubsection{Background Color Selection \& Preparation}
Our pilot studies suggested that the detectability or IR ink likely depends on the background color. Therefore, we included colors with varying RGB values, darkness ($K$) and of different luminescence to obtain a diverse set of colors. To cover a wide range of colors with different levels of darkness and luminescence, we first divided the darkness (calculated as an analog K value from RGB) into four equal 25\% intervals, ranging from 0\% to 100\%.
We then calculated the range of luminescence covered within each interval. As the darkness of a color increases, the luminescence range narrows, meaning the colors become darker. Therefore, we allocated four colors to each $\sim$25\% luminescence interval, starting with the lightest \new{colors in} Group 1, where darkness is between 0--25\%. Group 1 and Group 5 include the two extremes, no background color and black. Finally, we added four additional color hues to cover a large spectrum of real-world use cases and scenarios. 
\new{A detailed list of the colors tested in our experiment is provided in Table \ref{tab:color_coverage} in Appendix \ref{apedix: selection}}.

Finally, we printed the background colors with the IR ink gradient as described in \autoref{inks}. The size of the color bars was chosen to give participants a large enough area to visually inspect the color. As a result, we randomly split the 45 samples into groups of eight, resulting in six pages depicted in \autoref{fig:userStudyFigure}a.

\subsubsection{Design.}
We used a design within subjects in which each participant successively inspected each background color bar on all six pages. We fully counterbalanced the order of the six pages using a Latin square ($n = 6$).

\subsubsection{Participants.}
We recruited 18 right-handed participants (six females, twelves males), aged 18--31 (mean: 25.05; SD: 3.05) from the general public and the local university. Participants had a range of different educational and professional backgrounds, including computer science, education, cybersecurity, mechanical and electrical engineering, data science, and artificial intelligence. All participants reported normal/corrected-to-normal vision and did not report any known health issues which might impair their perception. Participants received a  \textdollar 10 Amazon voucher for taking part in the experiment. The study was approved by the MIT's Ethics Board.

\subsubsection{Experimental Protocol.}
Following a general introduction to the experiment, participants completed a demographic questionnaire. We then explained the task by showing them a set of prepared samples with background colors different from those used in the study. Next, we demonstrate how participants should explore the samples, i.e., moving the card from left to right with a slow and consistent speed. \new{They were instructed to sit comfortably with an approximate reading distance of 30--40~cm which mimics a typical real-world reading distance. We decided against fixating participants' heads, e.g., with a chin rest, to collect realistic real-world data}. The experimenter monitored that they stayed within a reasonable distance, i.e., not going too close to the sample. Once the participant indicated that there was a color difference, the experimenter obtained the corresponding IR ink value, which remained invisible to the participant. If a participant reached the end of the color bar and did not report a difference, the maximum IR ink value of 200\% was used as a response. Participants were given unlimited time to practice to ensure that they understood everything. They were encouraged to ask questions in case something was unclear. In total, the experiment took 35--45 minutes.

\begin{figure*}[t]
  \centering
  \includegraphics[width=1\linewidth]{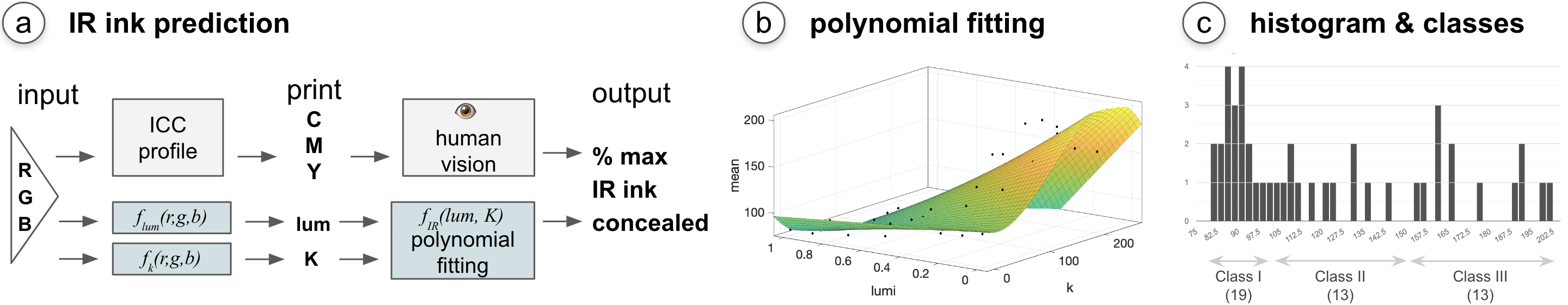}
  \caption{IR ink computation. (a) We attempt to mathematical model the limits of human perception to conceal IR ink. (b) We use polynomial fitting to estimate the IR capacity based on $lum$ and $K$. (c) The class limits were determined on the histogram of IR ink percentage.}
  \Description{.}
  \label{fig:IRInkComputation}
\end{figure*}

\subsubsection{Data Collection \& Analysis.}
We collected data from three sources: a pre-study demographic questionnaire; observations and field notes, and participants' responses to the 2AFC method. For each of the 45 background colors, we received 18 responses, resulting in a total of 810 data points. First, we performed outlier removal using the Box-plot method. Next, we statistically analyzed our data after verifying the parametric test assumptions at $\alpha$ = .05 using an RM ANOVA. In the presence of a main effect, we performed post hoc pairwise comparison t-tests adjusted using the Bonferroni-Holm \textit{p}-value adjustments.

\subsection{Results: Detectability of IR Ink}

\subsubsection{Printed IR Ink Can Remain Invisible to the Human Eye}
We compute an overall DT for each background color by averaging individual DTs of participants. The results can be seen in \autoref{fig:thresholds}, showing that for each background color, we obtained a DT greater than zero, reaching from DTs of 81\% (no background color) all the way to 200\% (black background color). This confirms \textbf{H.1} showing that the inkjet-printed IR content can remain invisible to the human eye. We also observed that the darker colors especially seem to suffer from high variance due to a wide range of participant responses. This may result from two reasons: (1) high-density IR ink has a ``greenish/grayish" appearance when printed. Background colors with the highest variance also show this color tone (see \autoref{fig:thresholds}) and, therefore, may be mistakenly perceived as IR ink. (2) we examined the high-variance samples after the experiment, and even though we fabricated high-quality prints, we noticed slight impurities. Given the nature of the task, i.e., we asked participants to fully focus on the sample and report a color difference as soon as they noticed it, we believe this might have pushed people to pick up on this, leading to even more conservative IR ink estimates.

\subsubsection{Maximum IR Ink Density Depends on Background Color}
\autoref{fig:thresholds} suggests that the invisible IR ink density depends on the background color. This is confirmed by our statistical analysis, which reveals the main effect of the background color on DT ($F(44) = 27.369$, $p < .001$, $\eta_{p}^{2} = .631$). 481/990 post-hoc pairwise comparisons were statistically significant after \textit{p}-adjustments, demonstrating that background color has a significant effect on the DTs in terms of the invisible IR ink density. Thus, we can also confirm \textbf{H.2}, that the maximum IR ink density depends on the background color.

\subsection{Predict Invisible IR Ink Density}
To facilitate the design process of watermark embeddings using invisible IR ink, we developed a system that estimates IR ink density for a given RGB color input (see \autoref{fig:IRInkComputation}). First, we performed two correlation analyses for $K$ and $lum$ for our 45 background colors on the amount of invisible IR ink to understand their influence. We found that $K$ shows a strong negative correlation ($\rho(43) = -.678$, $p = .012$) and $lum$ a strong positive correlation ($\rho(43) = .775$, $p < .001$). This means that increasing $K$ and decreasing $lum$ result in higher IR ink densities and vice versa. To give designers and creators a straightforward way to estimate invisible IR ink for any given RGB color input, we fitted a polynomial with 3 degrees of freedom, x = $K$, y = $lum$ and z = mean IR ink, using \textit{MATLAB}'s polynomial fitting function. The resulting linear model ($RMSE = 14.48$, $R^2 = 0.8477$) that can be seen in \autoref{eq:1} outputs an IR ink estimate based on the parameters $K$ and $lum$ given an RGB color input.


\begin{equation} \label{eq:1}
\begin{aligned}
f(K, lum) = & \, 113.7 - 0.1384*K - 25.43*lum + 0.0081*K^2 \\
            & - 0.5103*lum^2 + (0*K^3) - 0.0013*lum^3 
\end{aligned}
\end{equation}

Despite our efforts to include a variety of different colors, the amount of data that can reasonably be collected in psychophysical lab experiments without causing too much fatigue is limited. To improve the reliability of our method for colors that we did not test in our experiment, we first defined three classes based on the histogram of the mean IR ink.
\new{The limits of these classes were selected to represent the color distribution based on visual inspection of} \autoref{fig:IRInkComputation}c.
\new{Here, we notice that most tested colors are accumulated on the left side, i.e., representing lighter colors. The first class we define captures the accumulation of the lighter colors. In the histogram, the other two classes have the same number of colors to make the binning uniform.}

\begin{figure*}[]
  \centering
  \includegraphics[width=1\textwidth]{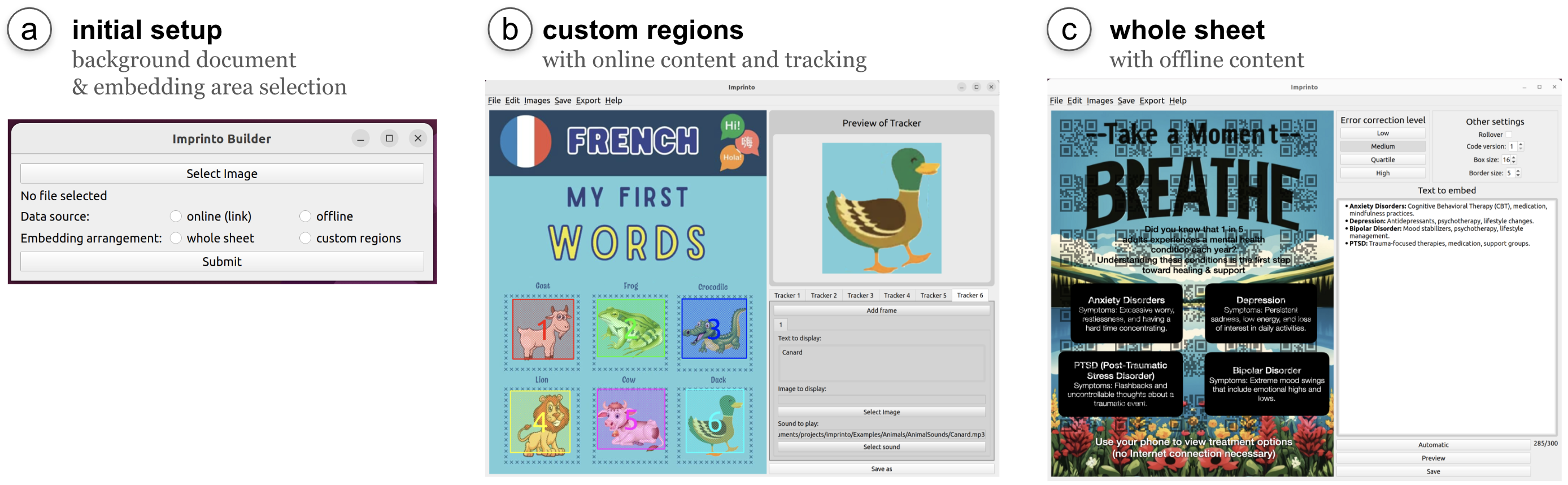}
  \caption{Our software tool for embedding \systemName~ watermarks. (a) The initial setup window is used to select between online/offline and whole/custom region modes. (b) Online mode allows users to designate custom regions for tracking and embeds invisible QR codes that instruct the app how to access the AR content from the internet. (c) Offline mode optimizes fitting to accommodate as much data on the page as possible.
 }
  \Description{.}
  \label{fig:UIexample}
\end{figure*}

\new{However, designers are free to modify this based on their use case, as our goal here is to provide an example for our applications using our methodology.} Class 1 (81\% to 101\%) contains the colors that allow the least density of IR ink. For all colors that fall into this category, following the polynomial function (1), the system recommends the very conservative value of 81\% IR ink (white paper). Similarly, for Class 2 (>=102\% to 154\%) it recommends 102\%, and 155\% for Class 3 (>=155\% to 200\%). Actual undetectability thresholds are most likely much higher. Still, we aimed to provide a method that allows designers to include any desired color while ensuring that users cannot notice the IR ink.

\subsection{Summary}
Our results show that the IR ink can remain \textit{invisible to humans regardless of background color} (\textbf{H.1}). Furthermore, we found that the maximum IR ink density depends on the properties, mainly the color $lum$ and $K$ of the background color (\textbf{H.2}). We developed a system taking RGB as an input, outputting a conservative lower bound for the invisible IR ink density given a background color. To this end, we provide evidence that IR ink can, in fact, remain invisible to humans; however, it is unclear whether inkjet prints with established IR ink densities can be detected using IR sensors. Therefore, the next section describes how we embedded, captured, and decoded our IR ink watermarks.

\section{Imprinto Watermarks}
In this section, we provide a comprehensive overview of the watermark embedding and detection process. The key steps involved are capturing the image, binarizing the image and extracting the QR code from the binarized image. Although extracting the QR code from an already binarized image can be done using commercially available software, the remaining necessary steps required the development of specialized modules.

\subsection{Embedding the Watermark}
We developed a software tool to embed markers based on the system and the invisible IR ink values determined in our experiment in Section~\ref{visibility-experiment}. To use our tool shown in \autoref{fig:UIexample}, users first select the document for which they wish to create embeddings. Then, they choose whether to utilize the whole sheet for embedding or only designated areas. In the latter, the user marks on the document where they want to embed content, and the system interactively generates layouts. Finally, users decide whether to rely on internet information (online) or create a fully self-contained (offline) document.
The tool then analyzes the RGB values of the background color of the specified image region. It determines the conservative lower bound of the invisible IR ink density using the underlying system described in Section~\ref{visibility-experiment}.

\paragraph{Online mode}
In the online mode, we take advantage of the ability to use the internet by hyperlinking multimedia content using QR codes.
In this menu (\autoref{fig:UIexample}b), the user can select the paper area that they wish to track using AR. An image score is computed for each selection using \textit{Google arcoreimg} tool\footnote{\url{https://developers.google.com/ar/develop/augmented-images/arcoreimg}} to ensure proper tracking and content overlay in AR.
In case the image is too difficult to track (i.e., the score is lower than 75), we inform the user and prompt them to choose a different area. Our tool allows users to associate each AR-tracked image with online multimedia content, such as text, image, and/or audio.
For each tracked image, the user can link a series ("frames") of these types of contents that users can loop through in AR. The underlying technology can be further generalized and used in order to incorporate complex multimedia content (such as videos or 3D models) or engineer further interactions between the objects. 


\paragraph{Offline mode}
In the offline mode (\autoref{fig:UIexample}c), we are currently limited to text-only content due to multimedia content requiring significantly more space.
However, this capacity that offline embedding offers can be used to embed hyperlinks or XML media, such as contact cards, or stylized and interactive text content via markup language tags of the user's choice.
Even though we offer an automatic analysis of the document and calculate a possible way of embedding the information, we further allow the user to fine-tune these settings (i.e., size, error correction level, and version of the QR codes) according to their specific requirements.

\subsection{Capturing the Watermark}

\new{\systemName~ media can be captured using devices with NIR image sensors. There are two main methods by which users can do this. First, more conventionally, users can point their devices at the physical media to extract \systemName~ content. This is particularly suitable for settings where the users are aware that the content is already embedded (e.g., an instructor informing students about \systemName~ verbally before handing out worksheets, or through a small visual cue on them; a book or magazine mentioning the AR functionality only on its cover---so not each page needs to carry a notice). 
Second, there are other types of devices that have constant scanning functionality~\cite{dogangun_rampa_2024, iyer_xr-penter_2025}. Emerging devices, such as smart glasses (see \textit{Meta Orion}\footnote{\url{https://about.meta.com/realitylabs/orion}}),  feature "always-on" IR scanning functionality to map the environment that we envision can be repurposed to detect human-invisible IR markers (see \autoref{fig:ApplicationsRobotic}). This, in turn, would enable the always-on and seamless capture of these markers without much physical effort. However, accessibility to the technology is not yet widely available, discussed in  
Section~\ref{NIR_AR_FormFactor}.} 

\begin{figure}[]
  \centering
  \includegraphics[width=0.47\textwidth]{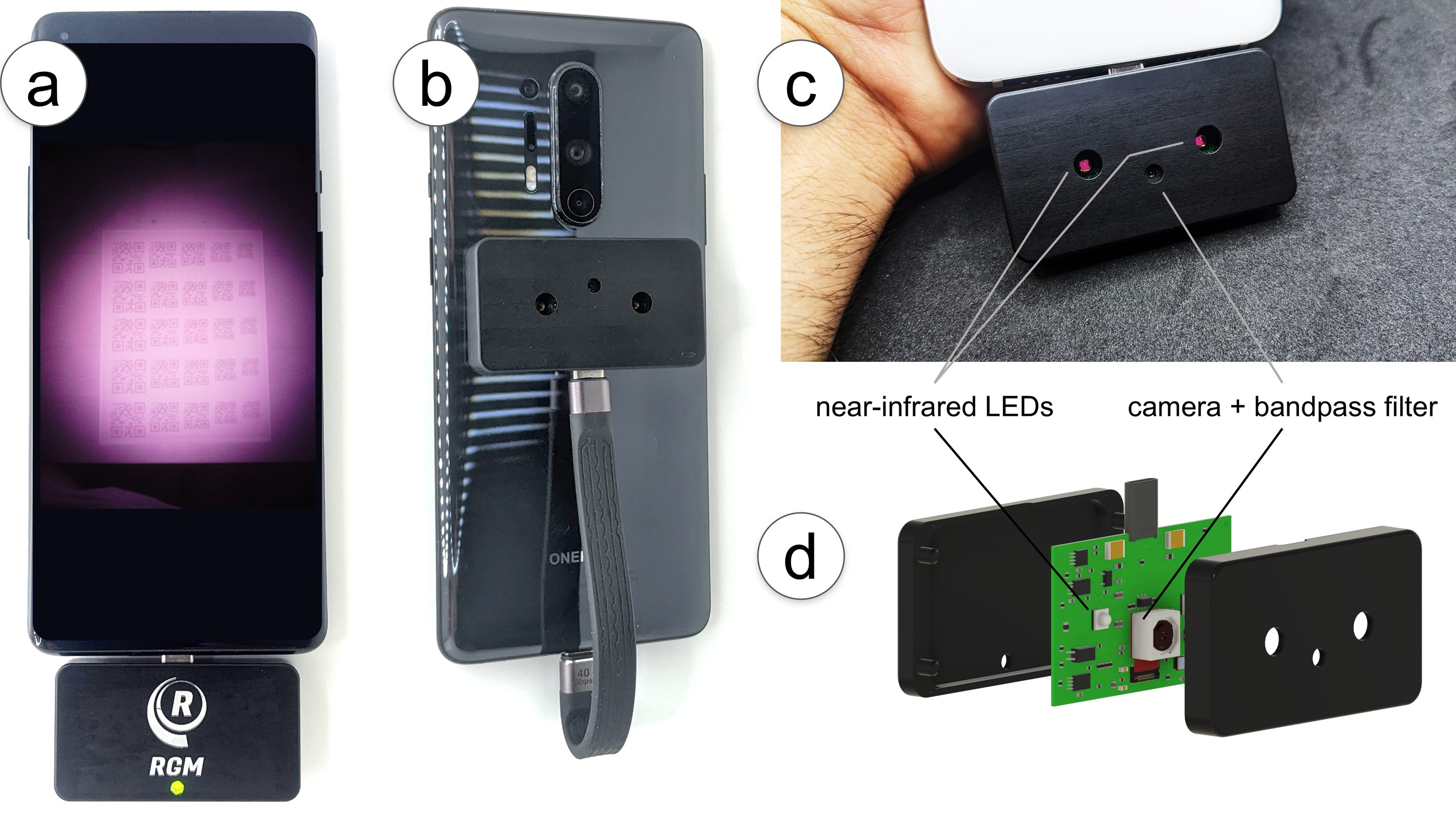}
  \caption{
  Our mobile imaging attachment comes with a 5MP sensor and two controllable NIR LEDs working at $850~nm$. (a) Module directly plugged into the mobile device. (b) Module coupled and lined with the back smartphone camera through a USB extender. (c) Front view of the module with the NIR LEDs powered up on a different device. (d) 3D exploded view, including the sensor, the LEDs mounted in the PCB, and the aluminum mechanical and electromagnetic enclosure.}
  \Description{.}
  \label{fig:RGMCameraModule}
\end{figure}

\new{To support this emerging technology, we developed a \textit{universal} \systemName~ detection and reader module for mobile devices.} 
The module was built to decode document watermarks based on the inkjet ink absorbance characteristics (Section~
\ref{spectrum}) by featuring a 5MP CMOS NIR camera and two controllable NIR LEDs, both operating at a wavelength of $850~nm$. The sensor module integrates a bandpass filter centered at the same wavelength. We designed and developed the PCB and the aluminum enclosure for mechanical and electromagnetic protection produced by \textit{Megacal}\footnote{\url{https://megacal.es/}}, with a size of $59.5\times33.0\times11.6~mm$. 
Figure~\ref{fig:RGMCameraModule} shows the components of the module.
Our proprietary module, \textit{RGMVision InfraRedCAM 1}\footnote{\url{https://www.rgmvision.com/product/rgmvision-infraredcam-1/}}, ensures universal mobile connectivity through \textit{USB 2.0} or \textit{3.0} and its \textit{USB-C} connector. 
As opposed to related works, this allows our device to offer universal compatibility, converting any mobile device, smartphone, laptop, or PC into an \systemName~ reader via its \textit{USB-C} support. Additionally, the two diffuse NIR LEDs (80°), controlled by two PWM signals, provide 24 different adjustable illumination levels for indoor applications.

We developed an \textit{Android} application (built for API level 26 and targeted for API level 30), based on \textit{Unity3D}, interfaces with the module through the \textit{Unity3D} plugin \textit{UVC4UnityAndroid v2.0}\footnote{\url{https://github.com/saki4510t/UVC4UnityAndroid}}.

\subsection{Decoding the Watermark}

We developed an ML pipeline to binarize the captured IR watermarks for robust detection, which we will make open-source. Because the IR camera captures the watermarks in a faint tone, off-the-shelf QR code readers are unable to recognize \systemName~ codes. Thus, we need an intermediary step where the contrast of the IR watermarks is increased via binarization, where the codes are converted into black-and-white patterns similar to conventional QR codes. Previous works~\cite{dogan_standarone_2023} have used traditional image processing filters for this step, whereas we use an ML-focused approach to increase robustness (Section~\ref{EvaluationDataEmbedding}). 
We next explain the individual steps that were necessary for building our pipeline, as well as how to use it in real time.

\subsubsection{Training the ML model}
\label{TrainingML}
\begin{figure*}[]
  \centering
  \includegraphics[width=0.9\linewidth]{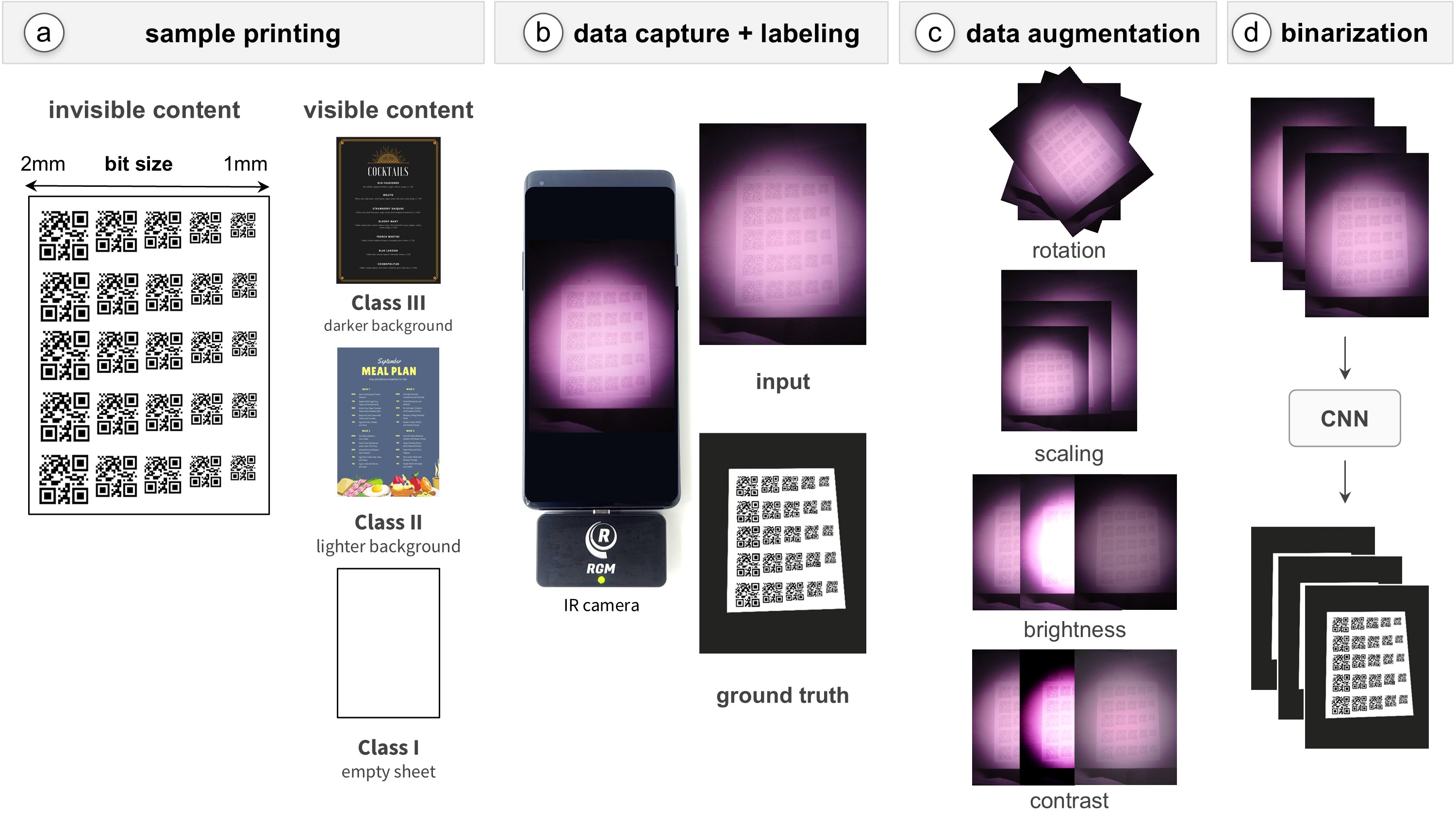}
  \caption{ML training. (a) Sample printing. (b) Data capture with our smartphone module and labeling processes. (c) Data augmentation. (d) CNN training for the binarization task.}
  \Description{.}
  \label{fig:IRTraining}
\end{figure*}

The training involved multiple steps, as shown in \autoref{fig:IRTraining}.

\paragraph{Data capture}

To ensure our ML model generalized to various conditions, we printed and captured multiple conditions with our setup. We intended to print different types of barcodes so the trained model can excel at binarization as an image task, regardless of the barcode type or content. Thus, we printed samples of three types of barcodes: a QR code with low ECC, a QR code with high ECC, and a Data Matrix code. Per sheet, we printed only one type of barcodes.

On each US letter-size sheet, we had a total of 25 codes that were created by repositioning and rescaling one of these initial barcodes in a grid-like pattern (see \autoref{fig:IRTraining}).
Varying the size of the code creates a more robust training set so that the model can recognize codes that appear smaller from a distance, at different angles, or with lower resolution.
We varied the the module (bit) size from $1~mm$ to $2~mm$, in increments of $0.25~mm$ from left to right to allow our model to generalize across different resolutions.

Each sheet was printed three times, once for each of the three possible IR ink quantity classes, and a matching visual for the CMY color content (darker background, lighter background, or blank), resulting in a total of nine sheets..
Regardless of the ink amounts and combinations, the model is expected to learn to detect and binarize the codes.
We then photographed each sheet using the IR camera from an identical distance of $25.5~cm$ in order to capture the letter-size sheet (see Section~\ref{DetectionDistanceCapacity}).

\paragraph{Ground truth labeling}

Using \textit{Photoshop}, we binarized the images in order to create a black-and-white image representing the ground truth. This process involved manually adjusting each code using various transformations (e.g., perspective warp) to match the pose of the code on the photographed sheet.

\paragraph{Data augmentation}

Using \textit{MATLAB}, we augmented our dataset by 500 images for each original image by performing a combination of geometric and color space transformations, including randomized changes in brightness, contrast, hue, and saturation.
\autoref{fig:IRTraining}c exemplifies the samples we generate in order to simulate different conditions that might occur due to varying physical conditions. 
Each image was rescaled to 450x600 pixels for efficient ML training.
In total, our resulting dataset had 4,500 images.
Such data augmentation in ML also helps us enhance model robustness and prevents overfitting by enlarging the dataset.

\paragraph{Training}
For the ML training, we employed a convolutional neural network (CNN) similar to \cite{dogan_sensicut_2021} but used the \textit{U-Net}~\cite{ronneberger_u-net_2015} architecture with a pre-trained \textit{ResNet-34}~\cite{he_deep_2015} as its backbone for black-and-white segmentation (i.e., binarization).
Using \textit{PyTorch}~\cite{paszke_pytorch_2019} and \textit{fast.ai}~\cite{howard_fastai_2020}, we utilized the \textit{Adam} optimizer~\cite{kingma_adam_2017}, and used cross-entropy as loss function.
We trained the model with a batch size of 8 for 30 epochs and for about 20 minutes each,  until training loss settled at 0.05. 
Because this is not a classification task but rather segmentation (binarization), we do not compute an accuracy percentage to evaluate the performance for the sake of the eventual code detection.
The binarized output will be passed onto a standard barcode reader, as reported in the next section.
We evaluate how the combination of our ML model and barcode reader enables data storage capabilities in Section~\ref{EvaluationDataEmbedding}.

\subsubsection{Real-Time Inference}
\label{RealTimeInference}

We developed a \textit{Unity3D} application that interfaces with both the phone's standard RGB camera and our IR camera module in order to both extract the embedded data and display it in a visually appealing manner by means of AR augmentation of the tracked environment.

The application works by constantly running the ML model on the feed from the IR camera and then redirecting the output to a code reader library \textit{Dynamsoft SDK}\footnote{\url{https://www.dynamsoft.com/}} for extraction. The output from the reader consists of a list of the values of all detected QR codes and their coordinates.
In the case of offline embedding, we concatenate all results from the read QR codes in order to reconstruct the encoded string. In the case of online embedding, each QR code corresponds to a location on the data server that allows the application to download further instructions.

\section{Technical Evaluation}
\label{EvaluationDataEmbedding}

\begin{figure*}[t]
  \centering
  \includegraphics[width=0.77\linewidth]{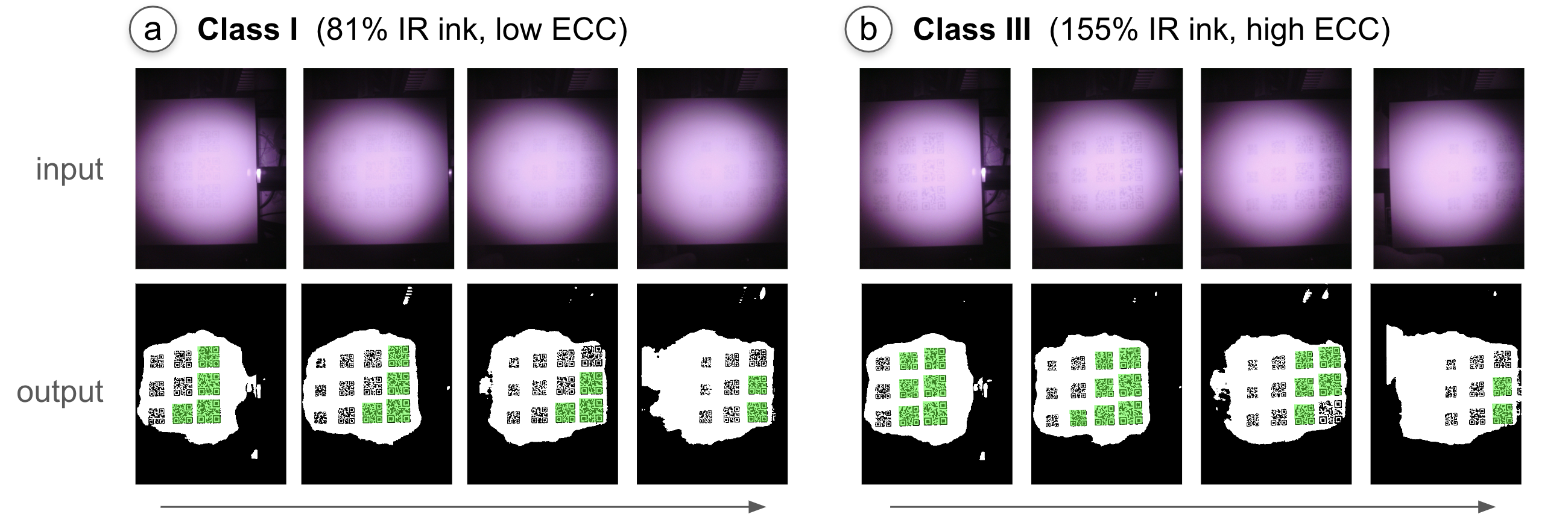}
  \caption{Examples of the samples used for bit size evaluation at $25.5~cm$ camera distance necessary for letter-size sheets.}
  \Description{.}
\label{fig:TechnicalEvaluationPreview}
\end{figure*}

In this section, we will evaluate various factors affecting the IR ink machine detectability and data capacity, such as IR ink density, detection distance and QR code's error correction code (ECC). In the following, we explain the effects of each variable. 
Starting with increasing the ECC, designers can improve the machine detectability of the QR codes, but increasing the ECC too much requires additional paper space for the same embedding.
On the other hand, by increasing the distance between the NIR camera and the IR content, the illuminance also decreases, and thus so does the machine detectability of the content.
Finally, by increasing the ink density, we increase the contrast, and thus also the machine detectability.
However, increasing the ink density might cause it to become visible to the human eye, and since we already determined the lower limits of human perception in Section~\ref{visibility-experiment}, the amount of IR ink that can be used in  \systemName~ documents is capped.

Therefore, we examine the effects of detection distance and ECC at invisible IR ink density levels on the data capacity of the sheet as it is crucial to better understand the trade-offs of using invisible IR content.
As Data Matrix codes do not provide the option to set an ECC, we focus on QR codes in our evaluation. Our results can guide design decisions when fabricating \systemName~ documents.

\subsection{Data Capacity Based on IR Ink Density, Detection Distance and ECC}
\label{DetectionDistanceCapacity}

\subsubsection{Procedure}
For the purposes of this evaluation, we will focus on the three invisible ink densities associated with the background color classes outlined earlier in this paper: 81\%, 102\% and 155\% (see \autoref{fig:IRTraining}a).
We focus on the distances necessary to capture a US letter-size sheet ($25.5~cm$ in order to capture an 8.5" x 11" sheet) and a US half letter-size sheet ($19.5~cm$ in order to capture a 5.5" x 8.5" sheet) as they present common real-world scenarios.
Please note that our current setup features an NIR camera that has a slightly larger field of view than the area illuminated by the integrated LEDs. However, for the purpose of evaluating the maximum capacity, we will assume that the whole sheet is uniformly illuminated.
After capturing the images, we use our image processing pipeline consisting of the ML model we trained (Section~\ref{TrainingML}) and the code reader (Section~\ref{RealTimeInference}) to evaluate the results.

\paragraph{Smallest Detectable Module}
In order to determine the maximum data capacity, we must assess the size of the smallest detectable module~\cite{dogan_standarone_2023}. Our sample printing precision was capped at 300 DPI, thus we were able to only assess the size of the smallest detectable module to within $0.085~mm$ (1" / 300 DPI). For each configuration of distance, ink density, and ECC (low, medium, and high), we printed a sample containing 12 randomly generated QR codes, that were not part of the training set used in Section~\ref{TrainingML}, ranging in total module size from $0.67~mm$ to $1.6~mm$ in increments of $0.085~mm$ arranged in a 3 by 4 grid (see \autoref{fig:TechnicalEvaluationPreview}). This range was determined based on empirical trials working with QR code detection after the initial training. In a real-world setting, each QR code would be exposed in multiple frames and from several slightly different positions.
In order to better reflect this in our evaluation, we aligned the camera with each of the four columns in the grid and captured a picture for each column.
We define the smallest detectable module as the smallest one that was detectable in at least one of the four pictures. Thus, our estimates for maximum data capacity only consider QR codes that can be detected. An example of the data for two such samples can be seen in \autoref{fig:TechnicalEvaluationPreview}.

\paragraph{Maximum Data Capacity}
Similarly to our perceptual experiment, our maximum data capacity estimates are conservative, since we enforce a quiet zone of at least four modules around each QR code in order to comply with the \textit{ISO/IEC 8859-1} standard. In this first iteration of our work, we assume that we cover the whole sheet with QR codes of identical size instead of trying a more complex packing algorithm that may involve mixing QR codes of different sizes. 
We included QR codes with varying data capacities, version 1 to version 7\footnote{\url{https://www.qrcode.com/}}, because in version 7 we reached our largest module size with a quiet zone of 4 modules.
Under the assumption that all QR codes are identical in size, we can iterate through all QR code versions (1--7) and try to fit them into a rectangular pattern to maximize the number of characters.

\subsubsection{Results}

\begin{figure*}[]
      \centering
      \includegraphics[width=0.89\textwidth]{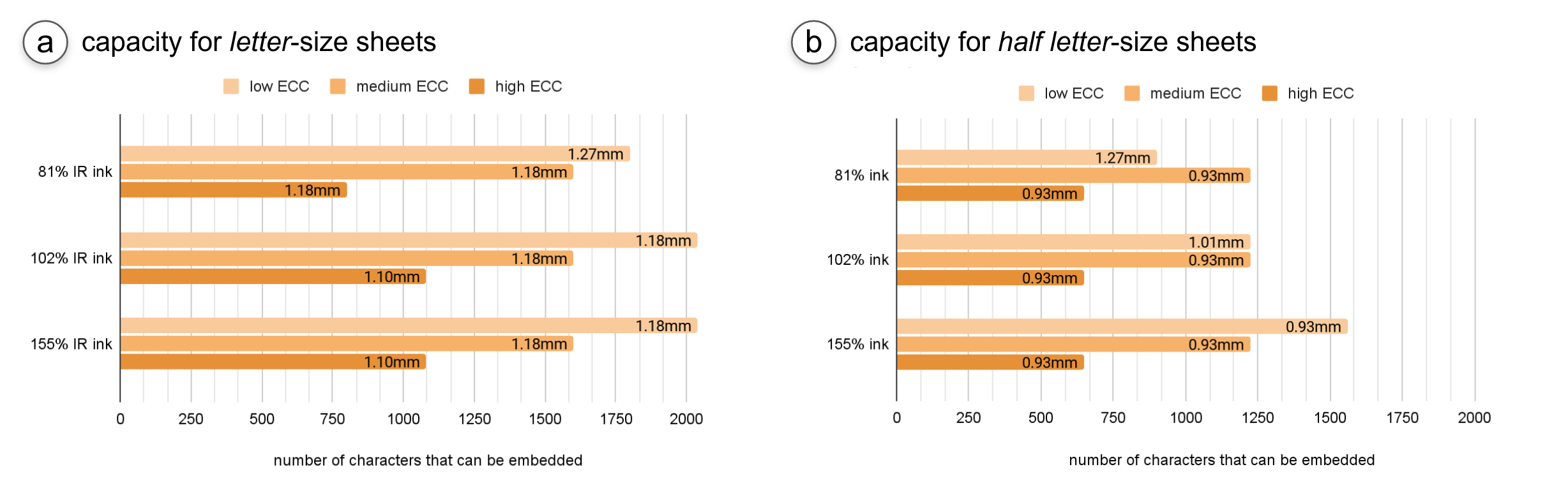}
      \caption{The effect of  IR ink amount and error correction code on the data capacity and module size of the sheet.}
      \label{fig:EvaluationCapacity}
  \end{figure*}

The key results of our data capacity evaluation can be found in \autoref{fig:EvaluationCapacity} and \autoref{tab:inkcomparison}~2.
Here, we report estimates for the binary encoding scheme of QR codes. In cases where the data is limited to numeric or alphanumeric characters, we can increase capacity by leveraging the QR code standard's native support.

\vspace{-0.1cm}
   
\paragraph{Letter-size}
The maximum capacity results for our letter-size sheets show that for all three IR ink densities, we achieve similar maximum capacities, with the smallest detectable module size of $1.18~mm$ under the medium and low ECC. 81\% IR ink seems to fall short of the other two classes with a smallest detectable module size of $1.27~mm$. This is to be expected since its contrast is rather low, making it the most challenging case. ECC showed the expected effect: a reduction in data capacity but improved detection, which can be seen as a further validation of our approach.  

\vspace{-0.1cm}

\paragraph{Half letter-size}
The findings for our half letter-size sheets suggest mixed results. First, the data capacities seem promising, given that only half of the space is available. The improved detection can most likely be attributed to the closer distance of the camera to the sheet. This is also reflected in the smallest detectable module size of $0.93~mm$, which is significantly smaller than in the letter-size condition. We want to highlight that we did not train our model described in the previous section on this distance, but it can still detect QR codes reliably. However, despite 155\%, the difference between low and medium ECC appears to have negligible effects.

\paragraph{Comparison to traditional binarization approaches}
To assess the effectiveness of the resulting CNN-based approach, we also compared it to traditional binarization methods used in previous work \cite{dogan_standarone_2023, dogan_demonstrating_2022}, which applies both contrast limited adaptive histogram equalization (CLAHE)~\cite{zuiderveld_contrast_1994} and adaptive thresholding~\cite{sauvola_adaptive_2000}, passing both binarized versions separately to the QR code reader.
We used the same dataset from the previous section to determine how effectively the CNN binarization can outperform the traditional filter-based image processing approach.
Our approach resulted in a 102.7\% increase in the total number of  QR codes detected compared to the previous method.
Notably, the CNN makes a larger impact on the detection of low-IR ink (Class 1) samples compared to high-IR ink (Class 3), i.e., an increase of 164.0\% vs. 92.7\%.
This is due to the fact that the pipeline for traditional methods vary based on factors such as lighting and the degree of code distortion, and may thus need constant adjustment of filter parameters.
In contrast, our CNN-based approach is more scalable, given its training dataset was crafted for our specific goal.

\vspace{0.2cm}

Together, our results provide users with estimates of how much content can be embedded using our IR ink watermarking technique.

\section{Use Cases and Future Applications}
In this section, we explore various applications of our watermarking technology designed for quick and integrated paper augmentation.
\new{We aim to demonstrate the breadth of use cases that \systemName~ can support and additional benefits that our method provides to users (e.g., invisible, mobile, privacy-preserving) across various contexts. Here, we present applications that are unique to \systemName, and we use replication for further validation \cite{Ledo2018Evaluation}.}

\subsection{Embedding "Paper DNA"}
\systemName~ could allow users to embed the digital metadata of the document as an intrinsic part of the physical paper. Utilizing this intrinsic "paper DNA," users could enable further authentication solutions.

\begin{figure*}[h]
  \centering
  \includegraphics[width=1.0\linewidth]{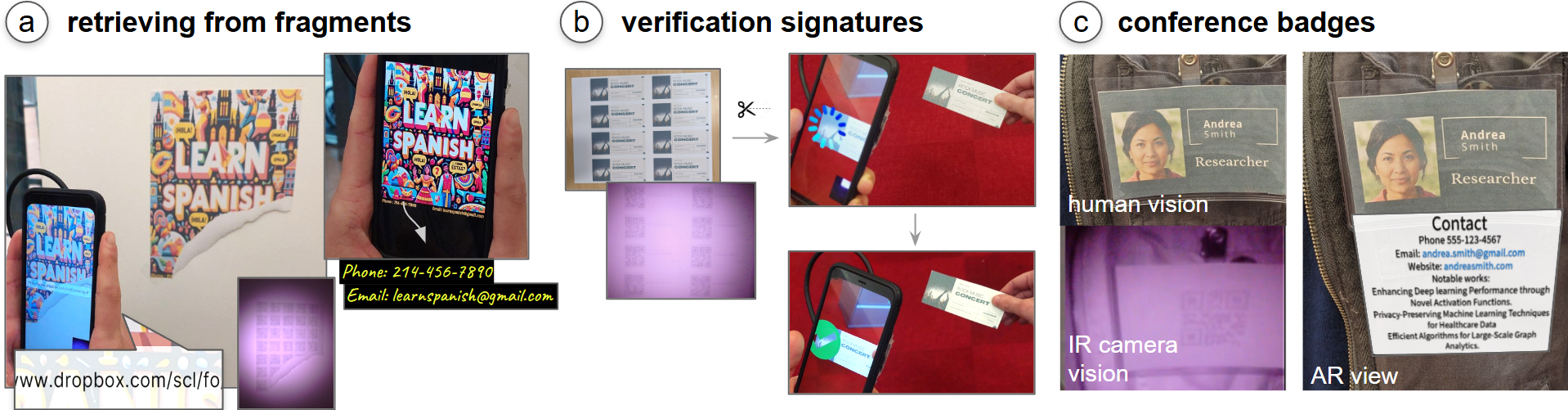}
  \caption{\systemName~ is used for making digital metadata an intrinsic part of the printed document to (a) recreate documents from fragments or (b) verify the authenticity of individual pieces. (c) The application augments normal vision by adding the AR content, providing additional information that would overload the limited space of a conference badge.}
  \Description{.}
  \label{fig:ApplicationsAR}
\end{figure*}

\subsubsection{Recreating Documents from Fragments}

The "Paper DNA" watermark allows fragments of documents to be scanned and linked back to the full digital copy. For example, if a poster for an educational event is partially torn, as shown in \autoref{fig:ApplicationsAR}a, a student can scan the remaining portion to retrieve the original poster and access the missing information. \new{The advantage of \systemName~ is that reconstruction can work even if the document is heavily fragmented. This is possible due to the high data capacity of a single marker that is sufficient to recreate the document. Because the information is repeatedly encoded without influencing the document's aesthetics, only one small part is necessary to retrieve its entire content.}
\newCameraReady{Maia et al. demonstrated this for 3D printed objects, where redundancy renders the information readable from a broken or damaged object to recover its original model
~\cite{maia_layercode_2019}, which inspired our application.}
\new{For paper materials, image-based tracking cannot easily achieve this because it would require training on fragmented documents, which does not scale well.}
We envision this feature can also benefit marketing strategies by including tearable sections in posters or ads with embedded links for future reference, \new{which are known to occupy too small space to include extra information, and \systemName~ can overcome these physical space constraints.}

\subsubsection{Imprinted Verification Signatures}
\systemName~ enhances document authenticity by embedding verification signatures in printed materials in combination with other techniques. For example, printed contracts could include invisible blockchain-based signatures or hyperlinks to the digital version for comparison against post-print modifications. If doubts arise, scanning with \systemName~ can confirm the document's originality. \new{The advantage is that this adds a second layer of "invisible" security by preventing people with malicious intent from reading names, passcodes, or signatures directly from the tickets, e.g., by shoulder surfing. This application exploits humans' unawareness of the presence of invisible IR content.}

\autoref{fig:ApplicationsAR}b shows an example where a friend group bought multiple tickets to a concert. The individual tickets, after being cut out from the sheet, each contain a verification code, which is scanned at the entrance of the venue.

\subsection{Augmenting Documents with Contextual Annotations and Multimedia}
\systemName~ seamlessly integrates digital interactivity into traditional paper media,  from educational enhancements to professional networking solutions.

\subsubsection{Smart Business Cards and Conference Badges}
\systemName~ could be used for fabricating \newCameraReady{intelligent business cards or conference badges as envisioned by Want et al.~\cite{want_bridging_1999}}, which could, for example, also enable access to restricted areas, digital contact exchanges, and enhance networking experiences at professional gatherings without any electronic components.
For instance, at conferences, attendees could wear badges and exchange business cards enhanced with \systemName, as shown in \autoref{fig:ApplicationsAR}c. Scanning a badge reveals the attendee's professional profile, contact information, and links to their work, facilitating networking without the need for physical contact or exchanging multiple items.
\new{This works best if attendees know that all badges or business cards support IR content--thus, they know how to retrieve this information. Additionally, they can hide information they do not want to share immediately with anyone who passes by because it is directly visible on the badge.}

\subsubsection{Interactive Flashcards for Learning}

\systemName~ enhances traditional flashcards by integrating multimedia, aiding tasks like language learning.
\newCameraReady{We replicate certain functionalities of CV-based flashcards, which include pronunciation guides via simple scans, turning materials into interactive experiences that improve outcomes \cite{ng2018treasure}.} 
However, our application also supports offline access to hints and answers, minimizing online distractions. Students can reveal hints or check answers through an app, promoting independent, self-paced learning. \new{Flashcards have very restricted space, severely limiting the kinds of tracking markers that can be embedded. \systemName~ uses the entire space of the flashcard, i.e., front and back, even without any content, because it does not interfere with it. Nevertheless, image or text detection could be used to achieve the same. However, we would expect \systemName~ to perform better during occlusive hand interactions with the flashcards since we can embed a larger number of evenly distributed markers.}

\subsubsection{Offline High-Capacity Data Access for Sensitive Topics}

\systemName~ can discreetly embed critical information into materials, useful for raising awareness about stigmatized topics or promoting sensitive products (e.g., healthcare) while ensuring privacy. For instance, a health organization can distribute posters on mental health with embedded AR markers, allowing access to additional information, such as treatment options, via a simple scan, as shown in \autoref{fig:teaser}c, without leaving a digital footprint. \new{This demonstrates \systemName's key contribution: enabling invisible, high-capacity offline data embedding without requiring users to constantly send data to third-party servers. We also see potential in using offline high-capacity data access to enhance the accessibility of printed media, such as alt-texts, to images for visually impaired people without adapting visual designs to accommodate markers.}

\begin{figure*}[t]
  \centering
  \includegraphics[width=\linewidth]{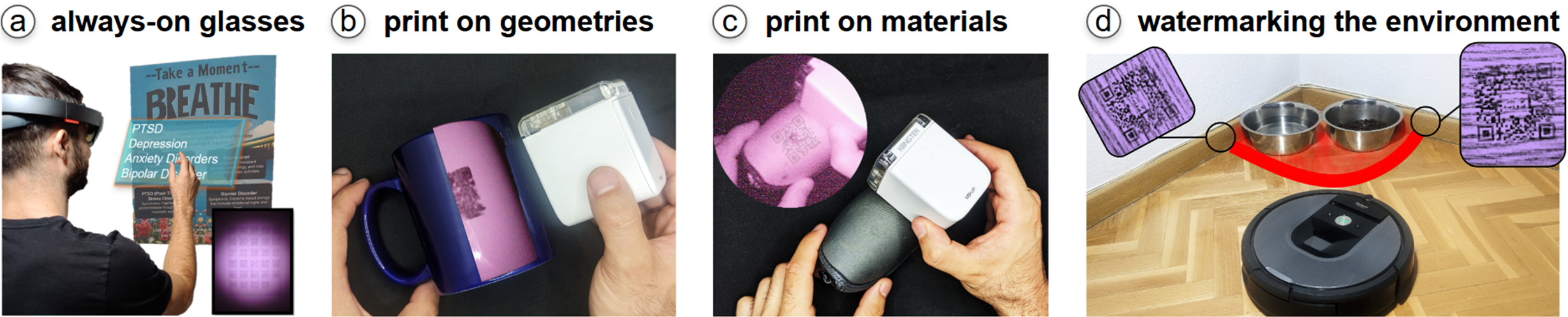}
  \caption{Envisioned extension of \systemName. \new{(a) A user wears AR smart glasses that capture \systemName~ content by exploiting the device's built-in always-on NIR cameras.} (b, c, d) Watermarking on different materials and surfaces to identify personal belongings and facilitate robotic navigation through invisible cues.}
  \Description{.}
  \label{fig:ApplicationsRobotic}
\end{figure*}

\vspace{0.2cm}
\new{\textbf{\systemName's Utility.} The examples in this section illustrate how IR watermarking can function as the "back end" for interactions with printed documents. Much like in web development or software engineering, this approach emphasizes a clear separation between the presentation layer (the "front end") and the data access layer (the "back end"). By abstracting and concealing technical, computer-readable components, our system offers a seamless, user-friendly interface while maintaining robust functionality. This separation enables new possibilities for interaction, surpassing the limitations of visual markers or standard QR codes in terms of abstraction and user experience. We argue that, if IR watermarks become the "back end" of printed media one day, it may pave the way for transformative applications without compromising the complexity and depth of the underlying technology.}

\subsection{Origami: Interactive Folding Instructions}
\systemName~ brings a twist to the traditional art of origami by embedding interactive, step-by-step folding instructions directly onto the paper. As users progress with their folds, new instructions appear on the exposed surfaces, as shown in \autoref{fig:teaser}b, based on the number and combination of the \systemName~ codes visible to the IR camera.
The advantage of \systemName~ is that even if the origami paper has no visible content (i.e., white sheet) in order not impact the look of the final artifact, the sheet can still carry invisible folding instructions.

\subsection{Sustainable On-Demand Watermarking}
\systemName~ empowers users to watermark existing posters and documents larger than letter-size format on demand (see \autoref{fig:teaser}a). This may help to safe resources by avoiding re-printing posters or documents in case of missing content. Traditionally, people often go through multiple iterations until they arrive at their desired design. \systemName~ enables people to focus on visual aesthetics with the possibility of adding content wherever and whenever needed.

\subsection{Beyond Paper Augmentation}
By using a mobile printer, \systemName~ could also add watermarks to plastic~\cite{ozdemir_speed-modulated_2024} (a), fabric~\cite{forman_defextiles_2020} (b), and wood~\cite{gui_draw2cut_2025} (c) and various form factors demonstrated in \autoref{fig:ApplicationsRobotic}. This is possible because the NIR ink is dye-based and can, therefore, be absorbed by these materials. In addition, our CNN detection pipeline can be fine-tuned for detecting, e.g., QR codes on convex or concave surfaces, highlighting the breadth of our contribution. This opens up new possibilities for embedding invisible identification markers to personal belongings (a) and (b). In \autoref{fig:ApplicationsRobotic}c, we also show how \systemName~ may be used for robotics use cases, in this case for robotic vacuum cleaners to safely avoid the pet's area.

\vspace{0.2cm}

Through these scenarios, \systemName~ shows its potential of integrating infrared inkjet watermarking into everyday interactions, and enabling solutions across educational, professional, and personal contexts.

\section{Discussion}

The value of hybrid paper-digital interfaces that augment physical documents with digital content has been well established~\cite{han_hybrid_2021}. To realize hybrid documents, we propose a novel watermarking method that uses IR-based printing to store digital content as an intrinsic part of the document. Unlike previous approaches that rely on a network connection and external storage for the digital assets, our approach with \systemName~ maintains the document format as the single container of both physical and digital contents. 
In this section, we discuss the limitations of using \systemName~ from the perspectives of authoring, printing, detection, and consumption. We also discuss future work to improve the approach.

\subsection{Towards Multimodal Content}
\systemName's interface currently supports only a rudimentary form of AR that only permits 2D visual content and audio and does not leverage multimodal content to its full potential. We hope that we will enable authors to embed more intricate objects and interactions in the future. Adding support for multimedia assets will pave the way to expanding on previously examined use cases ~\cite{alessandrini_audio-augmented_2014, rajaram_paper_2022}. The most impactful addition would be that of 3D assets since that would allow users to utilize the document as a spatial anchor.

\subsection{Effects on Printed IR Ink}

\new{Our experiments investigated human and machine detectability under controlled indoor lighting and viewing conditions.}
\newCameraReady{However, the perception and reliability of printed IR ink can vary significantly depending on external factors such as illumination levels, viewing angles, and the type of paper used.
One key limitation is that low-light conditions may reduce the effectiveness of human detectability due to insufficient IR reflection, while excessive illumination, such as direct sunlight, could lead to overexposure, affecting both machine readability and ink longevity. Future work could explore adaptive imaging techniques or optimized illumination setups to mitigate these issues. 
Regarding \textit{paper type}, we anticipate that glossy or coated paper may introduce reflection artifacts, which could interfere with machine-based recognition systems by creating specular highlights that obscure ink contrast.
Conversely, highly porous or rough-textured paper might cause ink diffusion, potentially reducing print sharpness and affecting detection accuracy.
A promising direction is to systematically evaluate these effects using an experimental framework similar to Xu et al.'s methodology, where QR code robustness was tested under varying lighting conditions, scanning angles, and environmental factors \cite{xu_art-up_2021}. Applying a similar protocol could help quantify how different conditions impact IR ink detection and visibility.}

As with any other dye-based ink, UV-dependent fading is an inherent limitation~\cite{maxmax_-_llewellyn_data_processing_ir_2022} that is important to consider for document reliability and permanence. Constant exposure to heavy UV sources such as the sun can cause the inks to fade over time.
\citet{willis_hideout_2013} showed how different IR ink types can be used in conjunction with UV-resistant coatings to achieve 98\% contrast preservation under office lighting conditions. We envision that for commercial use cases where long-term preservation is desired, a similar coating can be applied, which is sufficient for indoor applications.

\subsection{Invisible IR Ink}
We used a psychophysical experiment to determine conservative estimates for the IR ink densities that remain invisible to users for a wide range of different background colors. As with most experiments, we were limited by the number of background colors that we could include in the experiment. \new{Therefore, we do not know how our results generalize to other background colors, combinations of colors, and different graphic patterns. Nevertheless, we developed a system that facilitates embedding with invisible IR ink for any RGB color. Note that our method is very conservative and favors invisibility over machine detection. The actual DTs are likely much higher, which should further improve machine detectability. Nevertheless, we contribute a methodology that designers can apply to determine the "sweet spot" between visibility, data capacity, and machine detectability.}

We also want to highlight that the gradient IR ink and background color bars used in our psychophysical experiment differ from QR codes which may have had an effect on our estimated invisible IR ink DTs. We decided against using QR codes in the experiment because (1) it would have only been possible to test QR codes at predefined densities (e.g., at 20\%, 40\%, ...), limiting the precision of the study, and (2) the limited number of samples, which would not reflect the variability of QR codes of different sizes. As a result, our experiment also aims to provide general estimates for shapes other than QR codes and may, therefore, be used as a starting point for any type of invisible IR-printed marker.

\subsection{Discoverability and Practicality}
\label{NIR_AR_FormFactor}

\new{
To demonstrate \systemName, we used NIR-based fabrication and detection tools. While NIR cameras are getting popular in many handheld devices (e.g., \textit{iPhone} and \textit{iPad} use it for facial recognition and LIDAR 3D scanning), not all platforms currently give 3rd-party developers access to the raw NIR stream (e.g., only the processed depth map can be accessed on \textit{iOS}). We argue the interest in NIR applications will increase as more use cases are demonstrated by future projects.}

In our current implementation, \systemName~ documents can optionally be marked with a small icon or visual label on the document in our embedding tool
to allow users to discover embedded AR content.
\new{We envision that next-generation AR hardware can more fully leverage \systemName's utility. Compared to using a handheld device, always-on AR smart glasses such as \textit{Meta} \textit{Orion} could be constantly scanning the environment for hidden AR content~\cite{di_gioia_investigating_2022, campos_zamora_moirewidgets_2024}.
Currently, most AR glasses already leverage NIR cameras, but mainly for localizing the user and mapping their environment for 3D tracking purposes. With head-worn AR glasses with integrated NIR cameras used for \systemName~ sensing, the user would not have to manually capture objects using the phone during the interaction. We recommend future research to focus on implementing this on AR glasses.}

\section{Conclusion}

We introduced \systemName, a watermarking technique that embeds computer-readable information while remaining invisible to the human eye. Unlike previous methods, our approach utilizes the entire document, including white spaces, and works regardless of background color. Through a psychophysical experiment, we determined the maximum ink that can be embedded without being detected. We developed tools to support users in applying IR ink technology, including software for efficient information embedding and a universal camera module for capturing \systemName~watermarks. Our open-source ML pipeline processes these images for robust use with standard QR code readers. We demonstrated various use cases, highlighting the potential of invisible IR content for hybrid paper-digital interfaces and advancing watermarking techniques.

\bibliographystyle{ACM-Reference-Format}
\bibliography{ImmersiveDocs}

\onecolumn\appendix
\section{Appendix}\label{apedix: selection}
We provide a table for background color selection for the experiment and another table regarding the smallest module size and character capacity comparison.

\begin{table*}[h!]
\centering
\vspace{0.1cm}
\begin{tabular}{|c|c|c|c|}
\hline
\textbf{Group ID} & \textbf{Darkness}& \textbf{Calculated} & \textbf{Number of selected colors} \\
 & \textbf{{\small(analog $K$ value)}} & \textbf{luminescence ($lum$) range } & \textbf{{\small(roughly four colors per 25\% $lum$.)}} \\ \hline
1 & [0\%, 25\%] & (8.55\%, 100\%] & 16 \\ \hline
2 & [25\%, 50\%]& (5.7\%, 75\%] & 12 \\ \hline
3 & [50\%, 75\%] & (2.85\%, 50\%] & 8 \\ \hline
4 & [75\%, 100\%] & (0\%, 25\%] & 4 \\ \hline
5 & 100\% & 0\% & 1 \\ \hline
6 & \multicolumn{3}{|l|}{\small Four colors have been added to enhance color hue coverage.} \\ \hline
\end{tabular}
\vspace{0.1cm}
\caption[]{\new{Color selection table for the experiment. Colors for each tested group are chosen based on their darkness and luminescence ranges.}}\label{tab:color_coverage}
\vspace{-0.2cm}
\end{table*}\label{tab:studycolor}


\begin{table*}[h!]
\centering
\small\begin{tabular}{ccccc}
\toprule
 & \textbf{Letter size (distance $\geq$ 25.5cm)} &  & \textbf{Half letter size (distance $\geq$ 19.5cm)} & \\
\cline{2-3}\cline{4-5}
\textbf{Ink Type} & \textbf{Smallest size} & \textbf{Maximum capacity} & \textbf{Smallest size} & \textbf{Maximum capacity}\\
\midrule
Low EEC 81\% ink & 1.27mm & 1800 characters & 1.27mm & 900 characters\\
Medium EEC 81\% ink & 1.18mm & 1600 characters & 0.93mm & 1224 characters\\
High EEC 81\% ink & 1.18mm & 800 characters & 0.93mm & 648 characters\\
Low EEC 102\% ink & 1.18mm & 2040 characters & 1.01mm & 1224 characters\\
Medium EEC 102\% ink & 1.18mm & 1600 characters & 0.93mm & 1224 characters\\
High EEC 102\% ink & 1.10mm & 1080 characters & 0.93mm & 648 characters\\
Low EEC 155\% ink & 1.18mm & 2040 characters & 0.93mm & 1560 characters\\
Medium EEC 155\% ink & 1.18mm & 1600 characters & 0.93mm & 1224 characters\\
High EEC 155\% ink & 1.10mm & 1080 characters & 0.93mm & 648 characters\\
\bottomrule
\end{tabular}
\vspace{0.1cm}
\caption{Comparison of module sizes and capacities at different ink levels and distances.}
\end{table*}\label{tab:inkcomparison}

\end{document}